\input harvmac

\def\p{\partial}

\def\half{{1\over 2}}
\def\s{\sigma}
\def\sh{\hat{\sigma}}
\def\Ah{\hat{A}}
\def\Xh{\hat{X}}

\def\Dp{{D'}}
\def\At{\tilde{A}}
\def\Xt{\tilde{X}}

\def\a{\alpha}

\def\c{\gamma}

\def\e{\epsilon}

\def\s{\sigma}

\def\th{\theta}

\def\del{\partial}

\def\At{\tilde{A}}

\def \ss {{\s \hspace{-7.4pt} \slash}\;}
\def \ys {{y\kern-.5em / \kern.3em}}

\def\bd{
\begin{document}}
\def\ed{\end{document}}
\def\ft#1#2{{\textstyle{{\scriptstyle #1}\over {\scriptstyle #2}}}}
\def\fft#1#2{{#1 \over #2}}
\def\det{{\rm det\,}}
\def\tr{{\rm tr}}
\def\ra{\rightarrow}

\def\uha{{\hat {\underline{\a}} }}
\def\uhc{{\hat {\underline{\c}} }}

\def \Om {\Omega}
\def \bfd {{\bf d}}
\def \del {\partial}
\def \eps {\epsilon}
\def \Z {{\bf Z}}
\def \xb {\bar{x}}
\def \la {\langle}
\def \ra {\rangle}
\def \Omt {\tilde \Omega}
\def \la {\langle}
\def \ra {\rangle}


\Title{\vbox{\baselineskip12pt \hbox{hep-th/9911153}
\hbox{NEIP-99-017}}
}{\vbox{\centerline{
Matrix Theory in a Constant $C$ Field Background}}}

\centerline{Chong-Sun Chu$^{1}$, Pei-Ming Ho$^{2}$ and Miao
Li$^{3,2}$ } \centerline{$^{1}$ \it Institute of Physics}
\centerline{\it University of Neuch\^atel} \centerline{\it
CH-2000, Neuch\^atel, Switzerland}
\medskip
\centerline{$^{2}$\it Department of Physics} \centerline{\it
National Taiwan University} \centerline{\it Taipei 106, Taiwan}
\medskip
\centerline{$^{3}$\it Institute of Theoretical Physics}
\centerline{\it Academia Sinica, P.O. Box 2735} \centerline{\it
Beijing 100080}
\medskip
\centerline{\tt chong-sun.chu@iph.unine.ch}
\centerline{\tt pmho@phys.ntu.edu.tw}
\centerline{\tt mli@phys.ntu.edu.tw}

\bigskip

D0-branes moving in a constant antisymmetric $C$ field are found to
be described by quantum mechanics of the supersymmetric matrix
model with a similarity transformation.  Sometimes this similarity
transformation is singular or ill-defined and cannot be ignored.
As an example, when there are non-vanishing $C_{-ij}$ components,
we obtain the theory for D$p$-branes which is
effectively the noncommutative super Yang-Mills theory.
We also briefly discuss the effects of other non-vanishing
components such as $C_{+ij}$ and $C_{ijk}$.

\bigskip
\Date{November, 1999}

\nref\bfss{T. Banks, W. Fischler, S. Shenker and L. Susskind,
``M Theory As A Matrix Model: A Conjecture'', hep-th/9610043,
Phys. Rev. {\bf D55} (1997) 5112.}
\nref\bbs{For reviews see, T. Banks, ``Matrix Theory'',
hep-th/9710231, Nucl. Phys.  Proc. Suppl. {\bf 67} (1998) 180;
D. Bigatti and L. Susskind, ``Review of Matrix Theory'',
hep-th/9712072.}
\nref\cds{A. Connes, M. R. Douglas, and A. Schwarz,
``Noncommutative Geometry and Matrix Theory:
Compactification On Tori'', JHEP {\bf 9802:003} (1998),
hep-th/9711162.}
\nref\dh{M. R. Douglas and C. Hull, ``D-Branes And The
Noncommutative Torus'', JHEP {\bf 9802:008,1998}, hep-th/9711165.}
\nref\ch{C.-S. Chu and P.-M. Ho, Nucl. Phys. {\bf B550} (1999) 151,
hep-th/9812219; hep-th/9906192, to appear in Nucl. Phys. B;
V. Schomerus, JHEP {\bf 9906:030} (1999), hep-th/9903205;
F.~Ardalan, H.~Arfaei and M.M.~Sheikh-Jabbari,
JHEP {\bf 02}, 016 (1999) hep-th/981007.}
\nref\sw{N. Seiberg and E. Witten, ''String Theory and Noncommutative
Geometry'', hep-th/9908142.}
\nref\ss{A. Sen, ``D0-branes on $T^n$ and Matrix Theory'', hep-th/9709220;
N. Seiberg, ``Why is Matrix Model Correct'', hep-th/9710009,
Phys. Rev. Lett. {\bf 79} (1997) 3577.}
\nref\dhn{B. De Wit, J. Hoppe and H. Nicolai,
``On the Quantum Mechanics of Supermembranes'',
Nucl. Phys. {\bf B305} (1988), 545.}
\nref\dewit{B. De Wit, `` Supermembranes and Super Matrix
Models'', hep-th/9902051.}  
\nref\ak{T. Asakawa, I. Kishimoto, ``Comments on Gauge Equivalence
in Noncommutative Geometry'', hep-th/9909139.}
\nref\Ish{N. Ishibashi, ``A Relation between Commutative 
and Noncommutative Descriptions of D-branes  '', hep-th/9909176.}
\nref\corn{L. Cornalba, R. Shiappa,
``Matrix Theory Star Products from the Born-Infeld Action'',
hep-th/9907211;
L. Cornalba, ``D-brane Physics and Noncommutative
Yang-Mills Theory'', hep-th/9909081.}
\nref\bcd{D. Berenstein, R. Corrado, J. Distler,
``On the Moduli Spaces of M(atrix) Theory Compactifications'',
hep-th/9704087, Nucl. Phys. {\bf B503}, 239, (1997).}
\nref\tra{W. I. Taylor, M. Van Raamsdonk,
``Supergravity Currents and Linearized Interactions for
Matrix Theory Configurations with Fermionic Background'',
JHEP {\bf 04}, 013 (1999), hep-th/9812239.}
\nref\trb{W. I. Taylor, M. Van Raamsdonk,
``Multiple D0-branes in Weekly Curved Background'',
hep-th/9904095}
\nref\ikkt{N. Ishibashi, H. Kawai, Y. Kitazawa and A. Tsuchiya,
``A Large N Reduced Model as Superstring'', hep-th/9612115,
Nucl. Phys. {\bf B498} (1997) 467.}
\nref\miao{M. Li, ``Strings from IIB Matrices'', hep-th/9612222.}

\newsec{Introduction}

A background independent, nonperturbative M/string theory remains one
of the eminent problems in this ambitious program. There exists a
conjectured formulation of nonperturbative M theory in a flat
background, in the infinite momentum frame \bfss. This formulation
makes heavy use of intuitions from the D0-brane physics. As a first
step toward generalizing this formulation to a background independent
one, one may consider D0-branes on a curved background. However, it
turns out that D0-brane physics in this case resists a general
understanding.

When none of the maximal supersymmetry is broken, such as
compactifications on tori of dimensions less than 6, the problem
is not so difficult \bbs. In a somewhat seemingly simpler
situation where there is a general background of constant bosonic
fields, a formulation has not been proposed. One naturally divides
bosonic fields into two sets. The first consists of constant
metric. This problem is more or less trivial, since with a linear
coordinates transformation, the metric can be put into the
standard Minkowski form \foot{From a background independent
perspective, however, this is not a satisfactory solution, since
one always has to put the constant metric as parameters into the
Hamiltonian.
The $g_{-i}$ component in particular deserves more careful examination.
}. The second set consists of constant
antisymmetric tensor field  $C_{\mu\nu\rho}$.
In the presence of an interesting physical system
which the generalized matrix theory is supposed to describe,
one cannot always gauge away this constant background.
In the case when only $C_{-ij}$ are nonvanishing, and
$x^i,x^j$ etc. are compactified, there exists a proposal
by Connes, Douglas and Schwarz \cds . In this proposal, one replaces
the super Yang-Mills on the torus by the noncommutative super
Yang-Mills (NCSYM), with the noncommutative moduli given by
$\theta_{ij}=RC_{-ij}$, where $R$ is the radius of the
longitudinal circle. This proposal was later justified by
considerations in string theory \refs{\dh,\ch,\sw}, where the kind
of Sen-Seiberg's argument \ss\ in the decoupling limit is employed.

In Sec.2  we will start from the membrane action
with the coupling to the $C$ field, and discretize it to
obtain the D0-brane Hamiltonian. The resulting
quantum mechanics differs from the standard one only by a
similarity transformation. But this transformation may be
singular or ill-defined in various situations.
Two such circumstances are discussed. 
In particular, when one considers compactification or orbifolding,
different matrix models may result. 
As an example, in Sec.3 we will derive the NCYM
for a brane solution directly using this transformation.
We will show that the Connes-Douglas-Schwarz proposal
can be directly derived in a perturbation expansion
of the matrix theory without resorting to Sen-Seiberg's argument,
or to quantizing open strings at all.
We also show in Sec.4 how our similarity transformation
can be related to the map of Seiberg and Witten \sw\
between the noncommutative fields and the commutative ones.
Our approach is so general as to enable us to discuss
the effects of turning on other $C$ field components such as
$C_{ijk}$, $C_{+ij}$, $C_{+-i}$ in Sec.5.

Much remains to be done to unravel the physical effects of 
switching on other components of the constant C field, 
in different situations. The simplest is the effect of 
$C_{-ij}$ on D0-branes. Our discussion in the next section
indicates that there are effects even without compactification.
If matrix theory is correct, we expect that the spectrum of threshold
bound states of D0-branes is not changed. The first thing in
mind is then to calculate the Witten index again for the
system of N D0-branes.

\newsec{ A Similarity Transformation}

The Hamiltonian of multiple D0-branes can be 
derived by starting with the membrane action in the light-cone gauge, 
and replacing all physical variables, say $X^i(\s_1,\s_2)$, 
by matrices $X^i=\{X^i_{mn}\}$. 
Here we briefly review this procedure, leaving details to the original
literature \dhn. In the light-cone gauge, $X^+$ is identified with time
$\tau$, and $X^-$ becomes an auxiliary field satisfying the constraints
\eqn\cons{\p_a X^-+D_\tau X^i\p_aX^i+\hbox{fermionic terms}=0,}
where $\p_a=\p_{\s_a}$ and $D_\tau=\p_\tau+\{A_0,\cdot\} $ and the
Poisson bracket 
$\{A,B\} = \e^{ab} \del_a A \del_b B$
is defined with respect to the pair
$\{\s_1,\s_2\}$. We will concentrate on the bosonic variables,
since introduction of fermionic variables is straightforward.  The
Hamiltonian is written as
\eqn\ham{H=\int d^2\s \left( {P^+\over 2}(D_\tau X^i)^2
+{1\over 4 P^+} \{X^i,X^j\}^2\right).} To get to
the D0-brane Hamiltonian, we replace $P^+$ by $N/R$, the Poisson
bracket $\{,\}$ by ${N\over i} [,]$ ($1/N$ is treated as the
Planck constant), and the integral $\int d^2\s$ by ${1\over N} \tr$.
The D0-brane Hamiltonian thus obtained reads
\eqn\dham{H=\tr\left({1\over 2R}(D_\tau X^i)^2-
{R\over 4}[X^i,X^j]^2\right).}

We believe that the above procedure generalizes to the case when there
is a constant $C$ field. The coupling of the membrane to the $C$ field is
\eqn\cs{S_1= {1\over 6}\int C_{\mu\nu\rho}dX^\mu\wedge 
dX^\nu\wedge dX^\rho,}
and is a total derivative when all the components of $C$ are
constant. 
\foot{
We remark that our approach of treating the WZ term 
is different from that outlined in \refs{\cds,\dewit}.
In particular, ref.\dewit\ does not have an action
which is consistent with the Hamiltonian derived there.
}
Thus equations of motion as well as constraints derived from
the new action are the same as before. However, one cannot ignore
this total derivative term at the quantum mechanical level. For
instance, if we are to compute the quantum propagation of membrane
from a time $t_1$ to another time $t_2$, the propagator is given by
the path integral
\eqn\path{\langle \Psi (t_2)|\Psi (t_1)\rangle =
\int [DX]e^{i(S_0+S_1)},}
where $S_0$ is the membrane action without the $C$ field.  
Now since
$S_1$ is a total derivative, it can be written as two boundary terms
at time $t_1$ and $t_2$:
\eqn\bount{S_1= {1\over 6}\left(
\int d^2\s C_{\mu\nu\rho}X^\mu \{X^\nu,X^\rho\}(t_2)
-\int d^2\s C_{\mu\nu\rho}X^\mu\{X^\nu ,X^\rho\}(t_1)\right).}

The interpretation of the two boundary terms in the path integral is
straightforward: They simply ``renormalize'' the initial and final
wave functions. The new wave function 
$\hat{\Psi}$  then becomes
\eqn\renorm{\hat{\Psi}=U\Psi,}
with
\eqn\UC{ U=\exp\left({- i\over 6}\int d^2\s
C_{\mu\nu\rho}X^\mu\{X^\nu,X^\rho\}\right).}
For D0-branes, this
unitary operator $U$ is
\eqn\sim{U=\exp\left({-1\over 6}C_{\mu\nu\rho}
\tr X^\mu [X^\nu, X^\rho]\right).}
It is clear that our 
argument is independent of the precise form of the functional
integration measure in \path.

Notice that if
$X^-$ is involved in $U$, we should employ the constraints \cons,
thus the operator will contain the canonical momenta $P^i$,
and an ordering in the exponential in \sim\ must be chosen.
We shall discuss this in the next section.

The equivalent of ``renormalizing'' the wave functions is to perform a
similarity transformation on operators and to keep all wave functions
intact.  Given the operator ${\cal O}$, the new operator is 
$U^\dagger {\cal O}U$.  
If the similarity transformation operator $U$ behaves in a
reasonable way, the new theory obtained is identical to the original
one. However, if the similarity transformation is singular,
the new theory can be really a different theory. To see
that our similarity transformation is sometimes singular,
we consider two cases separately.

\vskip 5pt
\noindent{\bf The 1st Case:}

We first consider the case in which no $X^-$ is involved.
The exponential in $U$ is cubic in $X$. A simpler example of
this type is a single particle with a single coordinate.
For example, if  $$U=\exp(ia tx^2),$$ then a
time-independent wave function will
become time-dependent after this transformation,
or equivalently, the Hamiltonian will become time-dependent
in the Heisenberg picture. On the other hand, if
$$U=\exp (ia x^3),$$ then there is no effect at all.

For the case in which $X^-$ is involved, as we shall see in the
next section, the exponential in $U$ will take roughly the form
$X^3P$. Again consider the simpler case of a single particle with
$$U=\exp\left(-\left({a\over{n-1}}\right)x^n\p_x\right)$$
for $n>1$.
This operator is just $\exp(a\p_y)$,
where $y=x^{-(n-1)}$.
We start with a wave function $\Psi (y)$,
and demand it be normalizable and vanish at $x=\infty$.
Thus $\Psi(y=0)=0$. The transformed wave function is
$U\Psi(y)=\Psi(y+a)$. Its value at $x=\infty$ no longer vanishes,
instead it is $\Psi (a)$. Apparently this new wave function is no
longer normalizable.

These examples are quite similar to the case of
a charged particle on a circle 
when a constant gauge field is turned on. 
In this case the new wave
function is not periodic any more, if the original one is.  Similarly,
the above demonstrations show that the boundary conditions for wave
functions are changed under the action of $U$.

Although it would be very interesting to investigate further
such situations, in this paper we will only elaborate on
the following situation.

\vskip 5pt
\noindent{\bf The 2nd Case:}

Another situation in which the physics is changed by a similarity
transformation is when there are further constraints which
reduce the physical Hilbert space to a smaller space on which
the unitary operator is no longer well defined.
In the next section we will show that the noncommutativity
of D-brane worldvolume due to constant $B$ field background
can be understood in this way. Before we examine the D-brane case,
let us consider a toy model as a warm-up.

Consider a matrix model of $2\times 2$ matrices $X_i$ and a 
unitary transformation of the  matrix model by the 
operator $U=\exp\left(i\tr(\alpha\sum_i P_i)\right)$,
where $\alpha$ is a constant Hermitian matrix and $P_i$ is the 
conjugate momentum matrix of $X_i$. Obviously this unitary
transformation produces a shift $\a$ to all the matrices $X_i$: 
$(X_i)_{ab} \rightarrow \hat{X}_i=(X_i)_{ab} + \a_{ab}$  and 
doesn't change the commutation relations among the
matrix elements $(X_i)_{ab}$, $a,b=1,2$. Now suppose we are interested
in the commutative limit and impose 
the  constraints  $(X_i)_{12}=(X_i)_{21}=0$. 
The resulting $X_i$ satisfies $[X_i, X_j]=0$ and can
be viewed as a function on an ordinary commutative space consisting 
of two points. Obviously 
the constraint  kills some degree of freedoms of $X^i$
and the similarity transformation 
is no longer well defined in the constrained matrix
model. However one can also perform the similarity transformation
first and then impose the constraint, this way we obtain a new
matrix model  {\it different} from the original one, since
$[\hat{X}_i,\hat{X}_j]$ is now nonvanishing for generic $\a$.
In the next section, we will see that the constraints
effecting matrix model compactification is quite similar
in nature to the simple constraint we considered here.
It is therefore important to first perform the similarity
transformation and then impose the compactification constraints. 
This simple example illustrates  the same key
reason why the similarity transformation \sim\  for $C_{-ij}\neq 0$
results in the noncommutativity on a D-brane.

It should be clear from this example that this consideration can
be applied to orbifolds as well as compactifications.

\newsec{Noncommutative Yang-Mills from Similarity Transformation}

When the only non-vanishing components are $C_{-ij}$, we expect
that a brane solution in matrix theory is described by the NCSYM,
if all indices $i,j\dots$ are tangent to the brane. Similarly, if
$X^{i,j}$ are compactified, the NCSYM also emerges. The two cases
differ only in the Yang-Mills coupling, whose correct value can be
obtained by treating the operation $\tr$ properly in each case.
Thus, we shall not distinguish explicitly between the two.

Before going over to D0-branes, the $U$ operator \UC\ can be
rewritten as \eqn\minusx{U=\exp\left(- {i\over 2}\int d^2\s
C_{-ij}\{X^-,X^i\}X^j\right).}  Using the constraints \cons,
$$\{X^-,X^i\}=-D_\tau X^k\{X^k,X^i\}, $$
we now replace the Poisson bracket by a commutator,
the integral by trace, so the $U$ operator for D0-branes is
\eqn\dou{U=\exp\left({1\over 4}C_{-ij}
\tr\left[ [X^k,X^i],X^j\right]_{+}D_\tau X^k \right),}
where $[\cdot,\cdot]_+$ is the anti-commutator
and we have judiciously chosen an ordering in the trace.
Identifying the conjugate momenta
$P^{\mu}={1\over R}D_\tau X^{\mu}$,
we have the final form of the operator as
\eqn\foep{U=\exp(-I), \quad
I =  {1\over 4}\theta_{ij}
\tr\left[ [X^i,X^k],X^j\right]_+ P^k ,}
where $\theta_{ij}=RC_{-ij}$.
If we take the fermion part of constraints \cons\ into
account, $U$ will contain a part involving fermionic fields.
 
Everything we said so far is classical. When
we quantize the system, it is natural to adopt the Weyl ordering
prescription to have: $F(X) P \rightarrow
{1 \over 2} (F(X) P +P F(X) ) = F(X) P -i {\del \over \del X} F$ 
and the $I$ thus obtained will be
different  by an additional term of  
$I_{b}:={i\over 4} \theta_{ij} \tr[X^i, X^j]$. 
In fact, in order for $U$ to be a unitary operator,
it is necessary to use the Weyl ordering. However, $I_b$ 
can be nonvanishing only in the large $N$ limit and is 
proportional to the conserved membrane charge. It doesn't modify
the operator ${\cal O}$ at all and only
modifies the wavefunction by a phase. So long as we consider states
with the same membrane charge, these terms have no observable effects and
hence we will drop them from now on and use \foep\ as our definition.
But they cannot always be omitted if we consider
membrane processes with charge transfer.

The effect of adding the $C_{-ij}$ field background
in the matrix model is to replace every operator ${\cal O}$ by
$$ \hat{\cal O}=U^{\dagger}{\cal O}U. $$
With the anticipation that on compactification $X_j$ will be 
replaced by $i D_j = i \del_j + A_j$, 
we will split $X$ into $X^j= X^{j}_0+ X^{j}_1$ with $X^{j}_0$
corresponding to some constant 
background configuration that will be identified
with $i \del_j$ after compactification. 
$I$ is splitted correspondingly into $I= I_0 +I_1$, with
\eqn\IZero{\eqalign{I_0 &= -{i\over 4}\theta_{ij}\tr 
\left[ [X_0^i,X^\mu],X_0^j\right]_+
{\delta \over \delta X^{\mu}} \crcr
I_1 &=  -{i\over4} \theta_{ij} \tr
\left[ [X^i,X^\mu],X_1^j \right]_+
{\delta \over \delta X_\mu }
- {i\over 4}\theta_{ij}\tr
\left[ [X_1^i,X^{\mu}],X_0^j\right]_+
{\delta \over \delta X^{\mu}}.
}} 
One can separate the full operator $U$ into two parts 
\eqn\finalU{
U=U_1 U_0 \quad {\rm with} \quad U_0 = e^{-I_0},
}
and $U_1 = e^{-I_1}(1+{\cal O}(\th^2)) $.  
The omitted higher order terms ${\cal O}(\th^2)$
are terms that can arise in $U$ as $I_0$ and $I_1$ don't commute. 
The main reason for this separation \finalU\ with the
action of $U_0$ singled out explicitly
is that, roughly speaking, with $X_{0j}$ identified with $i \del_j$,
$U_0$ will result in the star product and
$U_1$ will relate the noncommutative $U(1)$ fields
to the commutative ones. 

Decompose $\hat{{\cal O}}$ into
\eqn\prop{ \hat{{\cal O}}=
U_0^\dagger{\tilde{\cal O}}U_0,\quad
{\tilde{\cal O}}=U_1^\dagger{\cal O}U_1. }
Let us first consider only the effect of $U_0$ and ignore $U_1$.
This can be viewed as the 0th order calculation in
a perturbative expansion in terms of $X^{\mu}_1$.
Denoting the matrix $X^{\mu}$
by $\Phi$ and using $P_{\mu}=-i{\delta\over{\delta X^{\mu}}}$,
we find
\eqn\seri{\eqalign{{\hat{\Phi}}=&\Phi +
({-i\over 2} \theta_{ij})[X_0^i,\Phi]X_0^j +\dots+ \crcr & +
{1\over n!}({-i\over 2}\theta_{i_1j_1})\cdots({-i\over 2}\theta_{i_nj_n})
[X_0^{i_1},\dots, [X_0^{i_n},\Phi]\dots]
X^{j_1}_0\dots X_0^{j_n} + \dots .}}

Now consider a solution representing a 
single dual D$p$-brane  whose longitudinal
directions coincide with those along which $\theta_{ij}$ is
non-vanishing. We use $\sigma_{i}$ to denote these directions.
The directions transverse to the brane will be denoted
by $X^a, X^b$. For the longitudinal directions, $X^i$ is replaced by
$iD_i=i(\p_i -iA_i)$ with $A_i$ being the $U(1)$ gauge field
and $\tr$ shall be replaced by $\int d \sigma$.
\foot{
For convenience, we will consider only $U(1)$ gauge symmetry on
the D-brane world volume in the following,
and its generalization to the $U(N)$ case is
straightforward. }
Let $X^i_0=i \del_i$, then
\eqn\propa{
\hat{\Phi}=U_0^{\dagger}\Phi U_0=
\Phi(\s_i+{i\over 2}\theta_{ij}\p_j)=\Phi(\hat{\s}_i), }
where by definition
$$f(\hat{\sigma})=\sum_{n=0}^{\infty}{1\over{n!}}({i\over 2}
\theta_{i_1 j_1})\cdots({i\over 2}\theta_{i_n j_n})
(\del_{i_1}\cdots\del_{i_n}f(\s))\del_{j_1}\cdots\del_{j_n}.$$
Thus we obtain a function whose arguments are
$\hat{\s}_i=\s_i+{i\over 2}\theta_{ij}\del_j$
and they no longer commute among themselves. It is
\eqn\nonc{[\hat{\s}_i,\hat{\s}_j]=i\theta_{ij}.}
In terms of the dual B-field on the D$p$-brane worldvolume,
$\theta={B\over{1+B^2}}$.
These are exactly the commutation relations
obtained by quantizing open strings on a D-brane.
Here we want to emphasize again that nowhere we have
resorted to string theory.

Note that our definition of the new function $\Phi (\hat{\s})$
through $\Phi(\s)$ is schematically
\eqn\defn{\Phi(\hat{\s})=\sum
{1\over n!}\p^n\Phi (\s)(\Delta \s)^n,}
where $\Delta\s=\hat{\s}-\s$ and is a derivative. On first sight, this
definition seems to be different from the usual Weyl ordering for
a function of noncommutative variables, which is
\eqn\weyl{\Phi(\hat{\s}_i)=\int dk
\tilde{\Phi}(k)e^{ik^i\hat{\s}_i}.}
As is well-known, the latter
definition obeys the star product. Our definition is instead
\eqn\oud{\Phi(\hat{\s}_i)= \int
dk\tilde{\Phi}(k)e^{ik^i\sigma_i}e^{-\half k^i\theta_{ij}\p_j}.}
But since $e^ Ae^B= e^{A+B}$ for $A,B$ commuting, the two
definitions are in fact identical. Therefore
\eqn\star{f(\hat{\sigma})g(\hat{\sigma})=(f*g)(\hat{\sigma})}
for any functions $f,g$ of $\tilde{X}_a$'s and $\tilde{A}_i$'s
with the star product defined by
\eqn\Moyal{
(f*g)(\sigma)=e^{{i\over 2}\theta_{ij}
{\del\over\del\sigma_i}{\del\over\del\sigma'_j}}
f(\sigma)g(\sigma')|_{\sigma'=\sigma} \; . 
}
It is now natural to interpret this as defining
the noncommutative algebra of functions
over a noncommutative space. It is satisfying to see that the
star product in noncommutative gauge theory has a simple origin
from matrix model.

It is convenient to introduce a left (resp. right) 
translational invariant ``vacuum''
denoted by $\rangle$ (resp. $\langle$)  which
is annihilated by all derivatives $\del_i$ acting from the left
(resp. right). 
So for instance
\eqn\vac{\Phi(\hat{\sigma})\rangle=\Phi(\sigma)\rangle. }
As a notation consistent with Stokes' theorem, integration on the
noncommutative space can be denoted as the ``vacuum expectation value''
$\langle\cdot\rangle$:
\foot{This notation is very natural in noncommutative geometry
for the following reason. In the quantum mechanics for a single
particle, a state can be denoted in terms of its wave function as
$\psi(x)\rangle$, and momentum acts on the state according to the
algebraic rules $[p,x]=-i$ and $p\rangle=0$. The inner product
of two states $\psi_i(x)\rangle$, denoted by
$\langle\psi_1^*(x)\psi_2(x)\rangle$, is just the integration of
$\psi_1^*(x)\psi_2(x)$.
}
$$\langle f(\hat{\sigma})\rangle=\int d^{p}\sigma f(\s),$$
where $\int d^{p}\sigma \; \cdot$
is the ordinary integration on a classical space,
and it should agree with the large $N$ limit of the trace $\tr$.

Here a puzzle arises. Consider two fields $\Phi_1(\s)$ and
$\Phi_2(\s)$ originally commuting with each other. After the similarity
transformation, $\hat{\Phi}_1$ and $\hat{ \Phi}_2$ do not commute,
while by naively applying \prop, one gets
$[\hat{\Phi}_1,\hat{\Phi}_2] =U^{\dagger}_0[\Phi_1,\Phi_2]U_0=0$.
The reason why this naive procedure is not correct
is as follows. When turning the matrices $\Phi_1$ $\Phi_2$ into
functions of $\s_i$, we have killed (infinitely) many degrees of
freedom since the two original large $N$ matrices
are not commuting in general. 
For instance one can first compactify the space on $T^p$
by imposing constraints like
\eqn\qc{V_i^{\dagger}X_j V_i=X_j+2\pi\delta_{ij}R_j,
\quad V_i^{\dagger}X_a V_i=X_a,}
and then let $R_j\rightarrow\infty$ in the end, if one wishes.
The similarity transformation
is ill-defined after the
constraints are imposed. This is just what happens in the warm-up
example in Sec.2. Instead, if we perform the similarity transformation
first and then impose the constraints for compactification (as we were
doing here in this section), we get the non-commutativity on the D-brane. 

In the above we have given discussions in the Hamiltonian
formulation. The Hamiltonian in the temporal gauge $A_0=0$ after
the similarity transformation becomes that of the noncommutative U(1)
gauge theory. If we want to recover the field $A_0$, for
consistency, it must be a noncommutative variable too.
Therefore the resulting action for the D$p$-brane is
obtained from the D0-brane Lagrangian
\eqn\Dp{ L =\tr\left({1\over 2}(D_0 X^{\mu})^2
-{1\over 4}[X^{\mu},X^{\nu}]^2\right)}
by replacing $X$ by $\hat{X}$,
usual product by star product
and $\tr(\cdot)$ by $\langle\cdot\rangle$.
Thus we obtain the well known NCYM Lagrangian.

\newsec{Relation to Seiberg-Witten Map}

In the previous section we have only showed the effect of
conjugation by $U_0$ on operators,
and NCYM is obtained as the 0th order approximation
of the exact matrix theory in $C_{-ij}$ background.
Now we consider the effect of conjugation by $U_1$
and examine the new fields $\tilde{A}_i$ and $\tilde{X}^a$.

Before we start, we mention that it is straightforward to 
repeat the idea of \sw\ to
derive the relation between the
noncommutative scalars and the usual scalars. Together with the result
for the noncommutative $U(1)$ gauge field, it is
\eqn\sws{\hat{X}^{sw}_a=X^{sw}_a-\theta_{kl}A^{sw}_k\p_l X^{sw}_a, }
\eqn\swa{\hat{A}^{sw}_i=A^{sw}_i-
\half\theta^{kl}A^{sw}_k(\p_l A^{sw}_i+ F^{sw}_{li}) }
up to first order in $\th$ for the gauge group $U(1)$.

For the action of $U_0$, the result \propa\ 
for a D$p$-brane solution is exact to all orders in $\theta$.
To the first order in $\th$,
the transformation by $U_1$ is given by
\eqn\Vone{ \tilde{X}_a= U_1^\dagger X_a U_1 =
X_a - \half \theta_{kl} A_k \del_l X_a+\cdots, }
\eqn\Vona{ \tilde{X}_i= U_1^\dagger X_i U_1 =
i[\del_i-{1\over 4}\theta_{kl}[(\del_i A_k),\del_l]_+ ]
+[A_i-\half \theta_{kl} A_k F_{li}] +\cdots. }

Ideally, the effect of $U_1$ should result in
the relations \sws\ , \swa\ of Seiberg and Witten which map the
commutative fields to the noncommutative ones.
To the lowest order in $\theta$, we should identify
our fields $\tilde{X}$ and $\tilde{A}$ with
Seiberg and Witten's noncommutative $U(1)$ fields
$\hat{X}^{sw}$ and $\hat{A}^{sw}$,
because they are multiplied with one another using the star product.
However, \Vone\ and \Vona\ are not exactly the same as
the Seiberg-Witten map, so we can not just identify
our $X$ and $A$ with their commutative $U(1)$ fields
$X^{sw}$ and $A^{sw}$.
To the first order in $\th$, the unwanted piece
(the second term in the first $[\cdot]$)
in \Vona\  can be absorbed in a change of coordinates. Let
\eqn\coord{ \s_k = \s'_k + \half \theta_{kl} A_l(\s'),  
} 
which implies a shift in $\del_i$
\eqn\coorddel{ \del_i'= \del_i -
{1 \over 4}\theta_{kl}[(\del_i A_k),\del_l]_+ +\cdots . }
This operator $\del'$ is chosen such that it is anti-Hermitian
and satisfies $[\del'_i, \s'_j]=\delta_{ij}$ and
$[\del'_i, \del'_j]=0$ (up to first order in $\theta$).
\foot{
Compared with the usual definition of derivatives after
a change of coordinates $\s\rightarrow \s'$,
\coorddel\ differs by the additional term
${1\over 4}\theta_{kl}\del_i\del_k A_l$.
This term can be accounted for by the change
of integration measure due to the Jacobian ${\del\s\over\del\s'}$.
} 
At the same time,
$$ A_i(\s)=A_i(\s'+{1\over 2}\theta A)
=A_i(\s')+{1\over 2}\theta_{kl}(\del_k A_i)A_l+\cdots,$$
and similarly for $X_a$,
giving exactly the extra pieces we were missing from
the Seiberg-Witten map in \Vone\ and \Vona .
Finally, we obtain
\eqn\ffa{ \Xh_a = \Xt_a (\sh'),
\quad {\rm with}\quad \Xt_a(\s')=
X_a(\s')-\th_{kl}A_k\del_l X_a+\cdots,}
\eqn\ffb{ \Ah_i = \At_i( \sh'),
\quad {\rm with}\quad \At_i(\s')=
A_i(\s')-\half\th_{kl}A_k(\del_l A_i + F_{li}) + \cdots,
}
where $\sh' = \s' +{i \over 2} \th \del'$.
These equations are exactly of the same form as the Seiberg-Witten map,
thus $\tilde{X}$, $\tilde{A}$
can be identified with the noncommutative $U(1)$ fields
$\hat{X}^{sw}$, $\hat{A}^{sw}$ in the Seiberg-Witten map
and $X$, $A$ with the commutative $U(1)$ fields.
To be exact, the separation of $U$ into $U_0$ and $U_1$
should be adjusted order by order in $\theta$
in order to reproduce the Seiberg-Witten map.

It is obviously more complicated to implement this kind of derivation
to higher orders in $\theta$, and one expects to meet ambiguities
if all we need is a map between the noncommutative and commutative
variables which preserves gauge transformations \ak .
On the other hand, it is also easy to see that in principle
the results at the first order in $\theta$ can be extended
to all orders by solving a particular differential equation
with respect to $\theta$. From $\hat{X}=U^{\dagger}XU$ and
$U=\exp(-\theta_{ij}J^{ij})$,
where $J_{ij}={1\over 4}\tr[[X^i, X_{\mu}],X^j]_+ P_{\mu}$,
one derives
\eqn\diff{\delta\hat{X}=\delta\theta_{ij}[J^{ij},\hat{X}]}
for a variation $\delta\theta_{ij}\propto \theta_{ij}$.
This is just the first order term in $\theta$
in the expansion of $\hat{X}$ in terms of $X$,
with all $X$'s replaced by $\hat{X}$.
Note that the derivation is valid only if $\theta_{ij}$
is varied by scaling. For a different path of variation
of $\theta$, the result is in general different \ak .
Note also that \diff\ is only analogous to the differential
equation in \sw ; they are different by a change of variables.

Note that we will not be able to perform the the change
of coordinates \coord\ for the case of $U(N)$ because
that requires the coordinates $\s$ to turn into a matrix.
The best one can do is to take the $U(1)$ part of $A$
for this change of coordinates. 
On the other hand, one should wonder why should one use
a single set of coordinates for all the $N$ D-branes,
while the background values of $B+F$ can be different
on each D-brane.
For the cases in which the expectation values of $B+F$
change significantly from brane to brane,
one expects that with only a single set of coordinates
satisfying $[\hat{\s_i},\hat{\s_j}]=i \theta_{ij}$, NCYM will not
be able to give a good description of the system for
any choice of $\theta$.
It is just because our formulation gives the exact result,
and the most general background of $B+F$ can be any $U(N)$ matrix,
that we are led to a situation where it seems natural
to introduce a matrix of coordinates.
A different consideration that seems to lead to the same conclusion is 
to start from a suitable
theory of Matrix open string and consider a sector of the theory which
describes $N$ coincident D-branes. Analogous to the situation in \ch,
it seems  natrual that the D-brane
worldvolume will  emerge  as  $U(N)$ matrices with
noncommutating matrix elements. We leave this possibility
for further studies.  

It would also be interesting to relate our results to recent
discussions on possible relations between noncommutative variables
and commutative variables \Ish\
\corn\ .

\newsec{Other Components of the $C$ Field}

In this section we consider the effects of other components of $C$.
Let us consider $C_{ijk}$ first. In this case the unitary operator $U$
is simply a function of $X^i$, so the $X$'s are not modified
by conjugation by $U$, but their conjugate momenta are changed.
It is easy to see that if the directions labeled by $i,j,k$
are compactified, $C_{ijk}$ will change the spectrum as:
\eqn\canon{P_i\rightarrow P_i-{i\over 2}C_{ijk}[X_j, X_k].}

It appears that without compactification, the above similarity 
transformation does not change physics, this is easy to see
by applying the transformation to the Hilbert space, rather than
to operators. An alternative argument is that $C_{ijk}$ are
not moduli in 11 dimensions. With compactification, the story
can change. If one dimension is compactified first, 
the matrix string results.
Let this dimension be $x^1$, then $C_{1ij}=B_{ij}$.
It is well-known in string theory that on a torus, $B_{ij}$ becomes
a genuine moduli, this implies that to get nontrivial physics we
need to compactify two more dimensions. We conclude that on $T^3$
parametrized by $(x^i,x^j,x^j)$, $C_{ijk}$ does have physical effects.
This is compatible with eq.\canon. Upon compactification, the
second term is proportional to $C_{ijk}F_{jk}$. The zero modes
of $F_{ij}$ will shift the canonical momentum $P_i$. If this
shift is not quantized (a vector on the momentum lattice), then
there is a net physical effect.
This effect was first discussed in \bcd\ based on physical
arguments.

The case of $C_{+ij}\neq 0$ is a little more interesting. Since in
the light-cone gauge $X^+=P^+\tau=(N/R)\tau$, we have
\eqn\esh{U=\exp\left(- {1\over 2R}C_{+ij}N\tr[X^i,X^j]\tau\right).}
It is nontrivial only when $\tr [X^i,X^j]\neq 0$. However $$iN\tr
[X^i,X^j]$$ is simply the membrane charge and is a conserved
quantity. The operator \esh\ simply shifts the energy by a
quantity proportional to the membrane charges.

If $C_{+-i}\neq 0$, both $X$ and $P$ will be modified by
the similarity transformation by a time-dependent piece. If we
compactify $X_i$, its momentum is quantized and the
physical effect of $C_{+-i}$ is manifested in the
momentum spectrum, in a way similar to the effect of a Wilson
line. But here the shift in spectrum is time-dependent.
It may be interesting to study in more details these cases.

The linear coupling of $C$ field in matrix theory
was also obtained in \tra\trb\ ,
it may be interesting to make connection of our formulation
with their approaches.

\newsec{Discussions}

We expect that our construction will shed light on the problem of
working out the complete Seiberg-Witten map.
It seems that serious progress in examining the AdS/CFT 
correspondence in a B field background
can be made only after the Seiberg-Witten map is well
understood. In general, one needs to find a general way to construct
local and gauge invariant operators. We leave this problem to future 
investigation.

The method of studying effects of $C$ field in matrix theory can be
readily generalized to the IIB matrix model \ikkt. There one starts with
the Schild action $S_0$ for a fundamental string. 
To include the $B$ field,
one adds a term to the Schild action
\eqn\iib{S=S_0+\int B_{ij}\{X^i,X^j\}d^2\sigma.}
Upon discretizing the world sheet, the second term becomes
\eqn\wzs{iB_{ij}\tr [X^i,X^j].}
To investigate the physical effects of this new term, one may
focus attention on a D-brane solution. It was already pointed out in
\miao\ that a D-string and in general a D-brane solution naturally
introduces the NCSYM as the effective world-volume theory. It would
be interesting to study this issue further by combining the term
\wzs\ with the ordinary IIB matrix action.

Come back to matrix theory. It is rather surprising to us that introducing
a constant $C$ background is implemented by a similarity transformation,
although this is imposed on us by the Chern-Simons term in the 
membrane action.
One may ask the deeper question in light of our observation: How does
a constant background emerges from a would-be background independent
formulation? In the past, adding background inevitably introduces new
degrees of freedom, this certainly is not in line with the idea of
a background independent formalism. A background should emerge as a 
collective solution of the existent degrees of freedom. Our work seems
to be a step further in this direction. We still put in the background
``by hand", but by merely reshuffling the old degrees of freedom through
a similarity transformation, instead of introducing new ones.

\vskip 20pt
\noindent
{\bf Acknowledgement }
\vskip 10pt

We thank Yong-Shi Wu for comments.
C.S.C. also thanks Adel Bilal for discussions and he 
is grateful to the Aspen Center of Physics,
the physics department of UC
Berkeley and Caltech for hospitality during various stages of the work.
He also likes to thank Michael Douglas for
comments on the previous version of this paper. 
The work of C.S.C. is supported by the Swiss National
Science Foundation.
The work of P.M.H. and M.L. is supported in part by
the National Science Council, Taiwan, and by
the Center for Theoretical Physics at NTU.
The work of M.L. is also supported by a ``hundred people project''
grant of Academia Sinica.

\listrefs
\end